\documentclass[11pt]{article}
\usepackage{amsmath}
\usepackage{amssymb}
\begin{document}
\title{Many-worlds interpretation of quantum theory, mesoscopic anthropic principle and biological evolution}
\author{A.Yu. Kamenshchik$^{1,2}$ and O.V. Teryaev$^{3}$}
\date{}
\vskip 5mm
\maketitle
\hspace{-0.8cm}
$^{1}$Dipartimento di Fisica e Astronomia and INFN, Via 
Irnerio 46, 40126 Bologna,Italy\\
$^{2}$L.D. Landau Institute for Theoretical Physics of the 
Russian Academy of Sciences, Kosygin str. 2, 
119334 Moscow, Russia\\
E-mail: kamenshchik@bo.infn.it\\
$^{3}$
Bogoliubov Laboratory of Theoretical Physics, Joint Institute for Nuclear Research, 141980 Dubna, Russia\\
E-mail: teryaev@theor.jinr.ru
\\
\begin{abstract}
We suggest to combine the Anthropic Principle with the Many-Worlds 
Interpretation of Quantum Theory. Realizing the multiplicity of worlds
it provides an opportunity of explanation of some important events which are assumed to be extremely improbable. The Mesoscopic Anthropic Principle suggested here is aimed to explain appearance of such events which are necessary for emergence of Life and Mind. 
It is complementary to the Cosmological Anthropic Principle explaining the fine tuning of fundamental constants. 
We briefly discuss various possible applications of the Mesoscopic Anthropic Principle including the Solar Eclipses and assembling of complex molecules.  Besides, we address the problem of Time's Arrow in the framework of the Many-Worlds Interpretation.    
We suggest the recipe for disentangling of  quantities defined by fundamental physical laws  and by an anthropic selection. The main emphasis is made on the problem of the biological evolution. 

\end{abstract}

\section{Introduction}
Since the days of its creation the problems of interpretation of quantum mechanics have been attracting attention of 
 persons working in this field as well as  of a more broad public, including philosophers, psychologists, biologists and even of people of art and literature (Jammer, 1974). The main feature of quantum mechanics, which distinguishes it from classical Newton mechanics is the fact, that even if one has a complete knowledge of a state of a system under consideration and would like to make a certain experiment, more that one 
alternative result of such an experiment is possible. The knowledge of the state of the system can permit us only to calculate the probabilities of different outcomes of the experiment, as was first understood by Max Born (1926).  
However, a natural question arises: how can we see only one outcome of an experiment and what happens with all other alternatives ?

The first attempt to answer this question was undertaken in the framework of the so called Copenhagen interpretation of quantum mechanics, which represents a collection of views and ideas of some of the founders of quantum mechanics. Especially important, at least from our point of view, were the contributions of Bohr, Heisenberg, Born and von Neumann, summarized in fundamental books (Bohr, 1958; Heisenberg, 1958, Born, 1955; von Neumann, 1955). 

The mentioned above scientists  emphasized different aspects of quantum theory and, correspondingly elaborated different features of the Copenhagen interpretation. Niels Bohr have suggested the notion of complementarity between different notions and approaches, which seemed to resolve apparent contradictions of the quantum theory. Besides, he insisted on the existence of the so called classical realm, where all the results of experiments and observations were registrated. Thus, the classical physics was considered not only as a limiting case of the quantum physics, but also as a pre-requisit of its very existence. 
Starting from his indeterminacy principle (Heisenberg, 1927) Heisenberg stated that the observable properties of microobjects arised due to the experiments and it is senseless to speak about their independent existence. Max Born insisted on the probablistic character of quantum theory. 

Finally, it was von Neumann, who formulated the mathematically rigorous idea of the reduction of the wave packet and in such a way had given a constructive picture of events, occuring in the process of quantum measurement. According to von Neumann, in the quantum mechanics coexisted two processes. One of them is a unitary deterministic evolution of the wave function, describing the quantum system, according to the Schr\"odinger equation. The second process takes place during quantum measurement and is called  the reduction of the wave function, when one of the possible outcomes is realized, while others disappear into  thin air. In the process of quantum measurement 
three players participate: an object, a measuring device and an observer, and the presence of the latter two corresponds in a way to the Bohr's 
classical realm. Thus, in this picture, everything which one calculates and predicts in quantum theory finds its explanation.

We would say, that the Copenhagen interpretation has played a very important role in the development of quantum mechanics and its applications. Perhaps, it did not teach the researchers how they should calculate the observable quantities, but it rather explained why these calculations can have some logical sense. Besides, it made the imagination of physicists more free and have made them more accustomed to the idea that the deterministic ideal of classical mechanics is not an absolute goal of the physical theory.   
However, some of the other founders of quantum theory such as Planck, Einstein, Schr\"odinger and de Broglie were not happy with the Copenhagen interpretation and thought that some rebirth of the classical   ideal was necessary. The most consistent attempt of such a rebirth was undertaken by D. Bohm and is known as de Broglie-Bohm interpretation (Bohm and Hilley, 1993).

On the other hand, the presence of two dynamical processes in the quantum theory looked logically unsatisfactory not only for researchers which could not leave aside the ideas of the classical determinism and it was here where the new epistemiological revolution was riping.
In 1957 a young researcher Hugh Everett has published a short version of his PhD thesis in the Reviews of Modern Physics (Everett, 1957). 
This paper had a rather modest title ``Relative-state formulation of quantum mechanics'' and contained a simple idea. We do not need the postulate of the reduction of the wave packet and hence, only one fundamental process exists in the quantum theory -- the unitary evolution 
governed by the Schr\"odinger equation. All the outcomes of the experiment co-exist and the objective result of the measurement under consideration is the establishment of correlations between the measured and measuring subsystems, which are treated on equal footing. 
Thus, there is no need in the special classical realm too. 

The Everett interpretation of quantum mechanics seemed to be quite logical and economical, however, this economy was achieved by means
of the acception of parallel existence of different outcomes of a quantum measurement and this was a critical point. Indeed, behind an 
innoncent mask of the relative states of two or more subsystems loomed a disturbing image of the co-existence of parallel worlds and of the splitting reality. Probably this fact explains rather a troubled history of entering of the Everett interpretation into the 
scientific community (Byrne, 2010). It is interesting that B. DeWitt was the referee of the paper in Reviews of Modern Physics and it took 
some time to him to accept its publication. Thirteen years later he has published the paper in Physics Today (DeWitt, 1970) and has edited a book (DeWitt and Graham, 1973), which have given a new birth to the Everett interpretation of quantum mechanics. Moreover, the name 
``many-worlds'' interpretation was also coined by DeWitt.  

For many years the many-worlds interpretation (MWI) of quantum mechanics was considered as something rather exotical by people working 
in quantum theory and its applications.  However, now the situation is changing due to two main developments in 
quantum physics: progress in the study of  quantum cosmology 
(see e.g. the book (Vilenkin, 2006)) and the birth of quantum informatics (Nielsen and Chuang, 2010). 
The quantum cosmology is a branch of the theoretical physics, which treats the universe as a unique quantum object. Thus, here there is no 
place for a classical realm, an external observer and other agents, which could be responsible for the presumed reduction of the wave function.
Therefore, growing interest to this field of research has aumented the interest to the Everett interpretation. On the other hand the 
quantum computational algorithms consider the parallel processes in a quantum system as a source of the drastic acceleration of the computing.
Thus, in some sense, the existence, at least on small scale, of ``parallel'' realities becomes a ``practical'' tool of quantum computational algorithms. 

Now, when both the researchers and the broad public become accustomed to notions of the many-worlds interpretation, one can note that, in general, the idea of existence of parallel realities is not so alien to a human intuition. It is enough to say that a lot of literature 
works, using directly the ideas of the many-worlds interpretation, have been created. We shall cite here only the book of science
 fiction 
with a smashing title ``The Coming of the Quantum Cats'' by Frederic Pohl (1986). 
However, it is more interesting that the idea of a real co-existence of different alternatives was expressed in literature without any connection with quantum mechanics and long before the creation of the many-worlds interpretation. A  reader, opening the collection of articles, edited by DeWitt and Graham (1973) finds there as an epigraph the citation from a story, written by Jorge Luis Borges,
``The Garden of Forking Paths'', written in 1941 (Borges, 1941). Much less known is a story, written by another Argentian writer, a friend 
of Borges, Adolfo Bioy Casares, which is called ``The Celestial Plot'', written even earlier in 1940 (Bioy Casares, 1940). The main character
of this story travels between parallel universes with the different histories, and these parallel universes look very much as Everett worlds.

As far as we know, almost nobody has done a juxtaposition of the many-worlds interpretation of quantum mechanics with a story, written 
in 1927 by a Russian Soviet writer Ryurik Ivnev ``An old man from Vladivistok'' (Ivnev, 1927). 
We shall permit ourselves to translate a couple of pieces from this story, used as epigraphs in the book by Barvinsky, Kamenshchik and Ponomariov (1988)).
`` The road, chosen by us, will be our main one, however, from here does not follow that other roads do not exist ?
They are real as well as that, chosen by us, and any of these roads would be our main road, if we had taken it.''
``What will give you my ``theory'', if it can be called in that way ? Now it gives you nothing. But in the future, provided 
 you have a curious intelligence, it would give you some pleasure. This pleasure consists in your feeling, that in your 
travel bag lay huge, and still not used, storages of the movies of your life.''

We have given these examples to argue that the main revolutionary idea of the many-words interpretation is not so counter-intuitive
and is quite compatible if not with a common sense, but at least with a general human fantasy.

Now, let us turn to the second topic of our paper.
The anthropic principle (AP) was proposed long ago (Dirac, 1937; Dirac, 1938; Dicke, 1961; Carter, 1970; Rosental, 1980;  Rosental, 1988;  Rosental, 1997; Barrow and Tipler, 1988), 
but recently it got a strong boost (Rubakov, 2006; Weinberg, 2005), 
connected with the development of cosmology (Garriga, Linde and Vilenkin, 2004; Pogosian, Vilenkin and Tegmark, 2006;
Garriga et al., 2006)
and string theory (Susskind, 2003).
The general idea of AP consists in the statement that existence of the (human) observer
imposes  important restrictions on the basic laws and fundamental physical constants.
As soon as these restrictions happen to be of tantamaunt importance, the required
values of physical constants appear to be extremely improbable.
This smallness of probability could be  compensated by the huge number of
universes constituting Multiverse. Under this term one should understand a complicated object which  may be formed by the process of the ramification of the spatial
structure of the universe due to the effects of spontaneous symmetry breaking producing inflationary expansion of the patches of spacetime.
Such an opportunity is inherent in the chaotic inflation models
(Linde, 1990).

Another source of multiversity is the existence of the so called string landscape which means that
the fundamental superstring theory contains  a huge amount of vacuum states, each of those
may lead to quite different universes with different physics.

Here we would like to elaborate our earlier suggestion (Kamenshchik and Teryaev, 2008) of MWI  as another source of
multiplicity opening the possibility of further extension of
applicability of AP. 
As soon as this multiplicity does not lead to the change of fundamental constants 
(although there are suggestions (Panov, 2011) to that the very early branchings may lead to the 
variations of these constants) 
we
are dealing with what we called "Mesoscopic" AP, corresponding to the scales intermediate between
cosmological and microscopic ones.
Let us emphasize that while the multiplicity coming from the string theory or from the inflationary cosmology 
is something still hypothetical, the multiplicity of the alternatives, which is present already in  quantum theory even for 
a relatively simple systems is something quite real. It is already here, and to recognize that it is so, it is enough to take the many-worlds interpretation of quantum theory seriosly. 
We have made an attempt to do it in the mentioned paper (Kamenshchik and Teryaev, 2008). Here we develop and extend our presentation. 

The structure of the paper is the following:
the second section is devoted to a brief review of the basic ideas of the many-worlds interpretation of quantum mechanics; in the third section we discuss branching of worlds understood in the sense of the
defactorization of the wave function and the problem of the preferred basis; in the fourth section
we consider the important  problem of irreversibility and appearance of the arrow of time in terms of
the many-worlds interpretation; the fifth section deals with the definition of the Mesoscopic
Anthropic Principle and its simplest applications to planetary systems; in the sixth section 
we discuss the relation between the Mesocscopic Anthropic Principle and the emergence of Life;
in the seventh section
we treat
the biological evolution in terms of variety of options provided by the quantum evolution; 
in the section eight we discuss possible relations between the problem of conscience and quantum theory;
in ninth section we try to argue that events, which have small probabilities of realization are no less important than those
having large probablities; 
in the last section
we discuss the main results and suggest some criteria for disentangling of  quantities defined by fundamental physical laws  and by an anthropic selection.

\section{Many-worlds interpretation of quantum mechanics}

As we have already mentioned, the many-worlds interpretation  of quantum mechanics  was suggested by H. Everett in 1957 (Everett, 1957)
and its invention was motivated by two factors. One of them was intensively discussed since the moment
of creation of quantum mechanics: it is the problem of reconciliation between two processes
present in the theory - dynamical evolution in  accordance with the Schr\"odinger equation and the
reduction of wave packet, responsible for an observation of the unique outcome of quantum measurement
when the quantum state represents a superposition of the corresponding eigenstates.
In the most popular Copenhagen interpretation of quantum mechanics such a coexistence of these
two processes was provided by the separation of the so called classical realm, which in some versions
was connected even with the presence of a conscious observer. Thus the desire of getting rid of the ambiguity
connected with the wave packet reduction postulate and having a unique quantum description of the Nature
stimulated the creation of MWI. In the framework of MWI the 
Schr\"odinger evolution is the only process,
 the principle of superposition is applicable to all the states including macroscopic ones and all the
outcomes of any measurement-like processes are realized simultaneously but in different ``parallel universes''. The very essence of the many-worlds interpretation can be expressed by one simple formula we are about to derive.
Let us consider the wave function of a system, containing two subsystems (say, an object and a device),
whose wave functions are respectively $|\Phi\rangle$ and $\Psi\rangle$ and let us the process of the interaction between these two subsystems is described by a unitary operator $\hat{U}$. The result of the action of this operator can be represented as
\begin{equation}
\hat{U}|\Phi\rangle_0 \Psi\rangle_{i} = |\Phi\rangle_i \Psi\rangle_{i}.
\label{unitary}
\end{equation}
Here the state $|\Psi\rangle_{i}$ is a quantum state of the object corresponding to a definite outcome of the
experiment, while $|\Phi\rangle_0$ is an initial state of the measuring device and $|\Phi\rangle_i$ describes the state of the measuring device, which has found the quantum object in the
state $|\Psi\rangle_i$. Now, let the initial state
of the object be described by a superposition of quantum states:
\begin{equation}
|\Psi\rangle = \sum_{i} c_i |\Psi\rangle_i.
\label{superposition}
\end{equation}
Than the superposition principle immediately leads to
\begin{equation}
\hat{U}|\Phi\rangle_0 \Psi\rangle = \hat{U}|\Phi\rangle_0\sum_{i} c_i |\Psi\rangle_i =
\sum_{i}c_i|\Phi\rangle_i \Psi\rangle_i.
\label{superposition1}
\end{equation}
The superposition  (\ref{superposition1}) contains more than one term, while one
sees only one outcome of measurement. The reduction of the wave packet postulate solves this puzzle by
introducing another process eliminating in a non-deterministic way all the terms in the right-hand side
of Eq. (\ref{superposition1}) but one. The MWI instead says that all the terms  of the superposition
are realized but in different universes.

The MWI looks the most consistent between interpretations of quantum theory, because it
ultimately reduces the number of postulates. Moreover, one of the proponents of MWI B.S. DeWitt says that in the framework of it
the mathematical formalism of the theory gives itself its interpretation (DeWitt, 1970).

Now, let us turn to the second motivation for MWI. In quantum cosmology there is no external observer and
hence, no, classical realm. Thus, MWI  matches quite well the quantum cosmology.

\section{Branching of Worlds and  the  preferred basis}

The opportunity to extract non-trivial physical consequences in the context of MWI is based on the
treating of the branching of worlds as an objective process. However, inevitable question arises:
decomposing the wave function of the universe one should choose a certain basis. The result of
the decomposition essentially depends on it. Thus, the so called  problem of the choice of the preferred basis arises 
( Deutsch, 1985; Markov and Mukhanov, 1988; Dieks, 1989; 
Ben Dov, 1990; Albrecht, 1992). The essence of the problem can be easily formulated considering
the same example of a quantum system consisting of two subsystems. Let us emphasize that now we would like
to undertake a consideration of a general case without particular reference to artificial measuring devices
and quantum objects (for a moment we consider this division of a system into subsystems as granted).
The only essential characteristics of the branching process is the defactorization of the wave function.
That means that if at the initial moment the wave function of the system under consideration was represented
by the direct product of the wave functions of the subsystems
\begin{equation}
|\Psi\rangle = |\phi\rangle |\chi\rangle
\label{direct}
\end{equation}
then after an interaction between the subsystems it becomes
\begin{equation}
\sum_{i} c_i |\phi\rangle_i |\chi\rangle_i,
\label{defact}
\end{equation}
where more than one coefficient $c_i$ is differentt from zero.
Apparently the decomposition (\ref{defact}) can be done in various manners. As soon as each term
is associated with a separate universe, the unique prescription for the construction of such a
superposition should be fixed.  We believe that the correct choice of the preferred basis is the  so
called Schmidt or bi-orthogonal basis. This basis is formed by eigenvectors of both the density matrices of the subsystems of the quantum system under consideration.
These density matrices are defined as
\begin{equation}
\hat{\rho}_{I} = Tr_{II}|\Psi\rangle \langle \Psi|,
\label{density}
\end{equation}
\begin{equation}
\hat{\rho}_{II} = Tr_{I}|\Psi\rangle \langle \Psi|.
\label{density1}
\end{equation}
Remarkably, the eigenvalues of the density matrices coincide and hence the number of non-zero eigenvalues is the same, in spite of the fact that the corresponding Hilbert spaces can be very different.
\begin{equation}
\hat{\rho}_{I} |\phi_n\rangle = \lambda_n |\phi_n\rangle,
\label{density2}
\end{equation}
\begin{equation}
\hat{\rho}_{II} |\chi_n\rangle = \lambda_n |\chi_n\rangle,
\label{density3}
\end{equation}
Consequently, the wave function is decomposed as
\begin{equation}
|\Psi\rangle = \sum{\alpha} \sqrt{\lambda_{n}}|\phi_n\rangle|\chi_n\rangle.
\label{density4}
\end{equation}

The bi-orthogonal basis
was first used at the dawn of quantum mechanics by E. Schr\"odinger (1935,1936)  for the study of correlations between
quantum systems and was applied to MWI in (Zeh, 1973; Barvinsky and Kamenshchik, 1990; Barvinsky and Kamenshchik, 1995a).
Recently, this basis is actively used for measuring of degree of entanglement, in particular, in relation
to quantum computing (Shimony, 1995). The expansion with respect to eigenvectors of spin density matrix
and density matrix positivity was also used in hadronic physics and non-perturbative QCD (Efremov and Teryaev, 1982;
Artru et al., 2009).

We believe that the bi-orthogonal basis being defined by the fixing of the decomposition of the system
into subsystems has a fundamental character and determines the worlds which result from the defactorization process. However, the subdivision of the system onto subsystems which implies the branching of the worlds
should satisfy some reasonable criteria which we are not ready to formalize at the moment (see, however the paper (Barvinsky and Kamenshchik, 1995b) for analysis of some relatively simple cases). One can say, that the decomposition into the subsystems should be such that the corresponding preferred basis were rather stable. For example, when one treats a quantum mechanical expreriment of the Stern-Gerlach type, it is natural to consider the measuring device and the atom as subsystems.

At the end of this section we would like to say some words about the comparison between 
the many-worlds approach to the interpretation of quantum mechanics and the decoherence approach. The decoherence approach to the problem of quantum measurement and to the problem of relations between quantum and classical properties of objects was proposed 
by Zeh in 1971 (Zeh, 1971). In the background of this approach lies an understanding 
of the fact that in any process of quantum measurement there are not two, but three participants: namely, not only the quantum object and measuring device, but also the rest 
of the universe - the so called environment. After the measurement, we can construct 
the reduced density matrix, describing the object and device, tracing out unobservable degrees of freedom of the environment. It appears that in many cases this reduced 
density matrix becomes quickly practically diagonal in some good basis, whose states are sometimes called ``pointer states'' (Zurek, 1981; Zurek, 1982) and behaves more or less classically. In such a way, the quantum state of the object and of the measuring device 
becomes a classical statistical mixture. The decoherence approach has a lot of useful applications, ranging from the description of properties of some molecules to quantum gravity and cosmology (see the book (Giulini et al., 1996)). However, from our point of view 
the decoherence approach to the problem of quantum measurement and to the problem 
of classical - quantum relations is less fundamental than the many-worlds approach. 

First, there is an essential difference between statistical principles in classical and quantum physics. In classical physics the probability is  ``the measure of our ignorance'' of the initial conditions or of the details of interaction while in quantum physics we cannot get rid of the probability even in principle, where is no analogue to the ``Laplace demon'', who can calculate everything. Thus, the transition to a classical statistical mixture does not resolve the problem of choice between different alternatives. 

Second, the decoherence properties of reduced density matrix depend crucially on the choice 
of the basis. Thus, the classicality is introduced into the theory already at the level of the choice of the basis. In the bi-orthogonal preferred basis approach, described above, the    
basis is defined by the chosen decomposition of the system under consideration into subsystems. After that, one can study the dynamics of different elements of the basis 
and to see if they behave classically (Barvinsky and Kamenshchik, 1995a). It appears, that 
sometimes classicality exists as a stable phenomenon, sometimes -- as a temporary phenomenon and sometimes it does not exist at all. Thus, the many-worlds interpretation,
insisting on  the primary role of the quantum theory with respect to the classical one,   
describes a more wide class of phenomena. Nevertheless, for a large class of situations,
the predictions of both approaches are close. It happens when the bi-orthogonal basis 
is close to the pointer basis.

\section{Time`s arrow}
The formalism of the many-world interpretation of quantum theory permits to reformulate the problem of a direction of time in a very transparent way. Indeed, the basic dynamics equations are invariant with respect to the operation of time reflection, while the macroscopic phenomena show the irreversibility or the presence of the arrow of time.  One of the quantitative manifestations of these phenomena is the growth of the von Neumann entropy (von Neumann, 1955)
\begin{equation}
S = -Tr (\hat{\rho}\ln\hat{\rho}) = -\sum_i \lambda_i \ln \lambda_i \equiv \sum_i S_i.
\label{entropy}
\end{equation}
where the last equality introduces, in the context of MWI  the notion of relative entropies of branches.
This entropy is minimal and equal to zero for a pure quantum state. Usually, the presence of the arrow of time is connected with the existence of some additional constraints on the solutions of fundamental equations. For example, choosing an initial state as a state with a low value of entropy, one naturally sees
its growth.
We make an observation that the branching process in the MWI naturally produces the states with a smaller initial  relative entropy (that is calculated by taking into account only one branch). In other words, after the measurement-like act of branching a new branch is in a factorized quantum state and the density matrices of
all its subsystems correspond to pure quantum states. This does not contradict to the increase of entropy
in the standard (Copenhagen) treatment of quantum measurement. In the latter case one is dealing after
the measurement with the classical statistical mixture of  various outcomes producing increase of entropy
which can be measured experimentally. At the same time in MWI the process of measurement (defactorization of the wave function) naturally implies the increase of entropy, but after the identification of an outcome of the measurement, when the defactorization of the wave function is completed, the relative entropy
(related to the branch where we live) becomes equal to $S_i$. Forgeting about other branches, which is equivalent to the reduction of the wave packet in the Copenhagen interpretation, corresponds to rescaling $\lambda \to 1$
and $S_i \to S^R; S^R(t_0)=0$, where $S_R$ is the redefined entropy after the branching happened at
time $t_0$.
Thus, relative entropy of each branch
is always growing, $S_i^R(t)> S_i^R(0)=0$,  so is $S_i$ and the usual measurable entropy of classical statistical mixture
which is just the sum (\ref{entropy}) of the entropies of the branches. Note that this nullification of relative entropy does not involve the distant regions of Universe which are the same for all the branches.

Thus, MWI provides another manifestation of the effect of boundary conditions which is present
in any explanation of irreversibility. The example of such boundary conditions
is, say, the correlations weakening in the Bogoliubov-Born-Green-Kirkwood-Yvon chain of equations leading to the
appearance of irreversibility. In another approach, when deriving (Zaslavsky, 2002)
the irreversible master equation
from the reversible Kolmogorov-Chapman equation it is sufficient (Teryaev, 2005) to assume the existence
of the initial conditions in the past.
The role of boundary effects for the
irreversibility  of field theory evolution equations implying the "scale arrow" , analogous to time's arrow, is discussed in (Teryaev, 2005; Artru et al., 2009). In turn, the irreversibility with respect to the time reflection
in field theory may appear
either because of T (or CP) violation at the fundamental level or because of its
simulation by imaginary phases of scattering amplitudes. The latter crucially depend
on the sign of $i \epsilon$ in the Feynman propagators which is imposed by the causal boundary conditions for Green functions. 
One may consider this as a sort of spontaneous symmetry breaking of time-reversal symmetry. 
This effect is giving rise to T-odd spin asymmetries (Efremov and Teryaev, 1985; Teryaev, 2000) being the subject of intensive theoretical and experimental studies.

In the actual case of MWI the choice of boundary conditions corresponds to the choice of factorized
wave function in the past, rather than in the future.
However, as MWI may be considered as "self-interpretation" of the mathematical formalism of quantum theory (DeWitt, 1970), the suggested approach may explain the fundamental phenomenon of Arrrow of Time in a similar manner.

To conclude this section we would like  to comment the differences between the many-worlds approach to the problem of the  irreversibility of time and the approach, developed by 
Prigogine and his school (Prigogine, 1980). The Brussel group suggested that the irreversibility 
should be introduced into both the classical and quantum physics already on the 
fundamental microscopic level. How  can one achieve this goal in spite of the fact 
that the corresponding dynamical equations are invariant with respect to the time inversion ?
The option of spontaneous breaking of T-invariance seems not to be considered by Brussel school.
It was suggested instead to  introduce an additional principle which prescribes the consideration of mixed states. In classical physics it means that one should consider some finite ``spots'' in the 
phase space instead of the points (pure states). In quantum physics it means that we 
should take into account only the states, whose density matrix is such that $\rho^2 \neq \rho$.
However, this requirement looks not so natural in classical mechanics. Indeed, 
if we consider a system in  a pure state all its subsystems will be also in pure states.
In quantum mechanics, it is quite common to have a system in a pure state and its subsystems in mixed states. Thus, the evolution of these subsystems generally is irreversible.
In terms of the many-worlds interpretation it means that the branching of the Everett worlds 
is generally an irreversible process. Let us conclude this discussion by mentioninng
that MWI makes the spot, like  any 
statistical ensemble, the very natural physical notion.

\section{Planetary Coincidences and Mesoscopic Anthropic Principle}

It is usually believed that the suitable values of fundamental constants are sufficient
for emergence of stars, planetary systems and all the astrophysical objects required for
apperance of life. However, there are a number of observations pointing to the special, privileged, role
of the Solar system (see e.g. (Gonzalez and Richards, 2004)). All the values
describing this privileged position
cannot involve the fine-tuning of neither constants of elementary particle physics nor cosmology.
Therefore we call such a coincidences the mesoscopic anthropic coincidences and
the related selection the Mesoscopic Anthropic Principle (MAP).

Note that understanding of Earth exceptional properties  developed relatively recently and may be considered as a discovery of present Millenium.
The famous Polish thinker, philosopher and Science fiction writer, Stanislaw Lem was writing (Lem, 2001) in an essay summarizing the outcomes of his numerous predictions
``Last time I was quite impressed by reading of the solid American works published in 2000, discussing the extreme uniqueness of our planet with its biosphere, and, indirectly but expressively the authors convince us that all of us
(together with fellow  yeast) are the only living creatures in all the galaxy called Milky Way''.
 
Note also, that the notion of MAP in its application to Earth specifics is very close to ``Planetary version'' of AP
suggested by R. Dawkins in his bestseller (Dawkins, 2009)-pp.62-69, whose 1st edition, appeared in 2006, 
unfortunately, was not known to us when the e-print version of (Kamenshchik and Teryaev, 2008) was released in 2007. 

The first natural opportunity to find the privileged values of planetary characteristics
is to explore the vast number of galaxies, stars, and planets in our Universe (Muller, 2001; Dawkins, 2009).
Note, that the necessity of this large number provides a sort of answer for one line
of possible criticism of AP suggesting that the existence of such a
large Universe is hardly necessary for the life on the Earth, this argument being
best expressed by S. Hawking who was saying that 
``our Solar system is certainly a prerequisite for our existence... But there does not seem any necessity for other galaxies to exist'' (Hawking, 1993).

At the same time, the selection among the large number of distant astrophysical objects
does not seem sufficient if some fine-tuned value of mesoscopic parameter is required.
For this aim the small changes of the relevant parameter within the required range
are important. This is exactly what happens in the chaotic inflation or stringy landscape
and allows for a fine tuning of fundamental constants
\footnote{Such a small changes of some parameter constitute, in fact, the cornerstone of Darwinian natural selection, see also the next sections}. At the same time, the emergence of supportive values of the planetary characteristics 
due to  small variations of initial values of Universe evolutions seem unnatural due to the very large time interval and likelihood of instabilities and chaos of the respective evolution. 

As a possible solution of this problem we suggest that the MWI is a source of
small variations of mesoscopic planetary constants in different worlds occurring at various time scales.
We assume that the measurement-like quantum interactions leading to the branching
occur all the time independently of the presence of (conscious) observer and produce
the planetary systems in parallel Everett worlds whose parameters differ by small amount.

The example of planetary fine-tuning is provided by Solar eclipses requiring the
coincidences of angular sizes of Sun and Moon, as seen from the Earth.
There is currently no explanation of this coincidence, apart from teleological arguments. 
At the same time, this coincidence would be explained if the
eclipse were necessary for some stage of the emergence of life ( Teryaev,unpublished; 
Kamenshchik and Teryaev, 2008). This
does not seem completely impossible, although there is no evidences in favour of such
a relation. One possibility is the emergence of life due to photochemical reaction requiring
the shadowing of strong ultraviolet radiation of the Sun but presence of the radiation of Solar
Corona. 
Another possible role of the Moon having the angular size similar to that of Sun can be due to tidal force resonance (Vilenkin, 2010). Note also that the discrepancy of sizes due to a change of distance between Moon and Earth may be a signal that the Moon played its role not so long ago.   


\section{Mesoscopic Anthropic Principle and Emergence of Life}

However, even a suitable planetary environment does not lead automatically  to the  emergence of the primitive life.
In fact, the probability of emergence of the first DNA molecule to start the  simplest replication cycle is about 
$10^{-400}$ (Chernavsky, 2009), while the number of attempts (realizing the famous Darwin's idea of ``small warm pond'') is about  $10^{-29}$. Let us note that this order of magnitude difference in the argument of exponential will persist in the popular model of RNA world.   

This huge number may be compensated by a huge number of attempts in the framework of MWI.
The crudest estimate (Kamenshchik and Teryaev, 2008) of the number of the Everett worlds
produced up to the present moment was based on the assumption that it is the Planck constant $\hbar$ which selects the
measurement-like interactions leading to defactorization.
For dimensional reasons when determining the number of worlds
it should be divided by some constant with the dimension of action or phase space, characterizing the whole Universe.
The relevance of the scale of the whole Universe can be seen from EPR measurement of two particles with arbitrarily large separation between them.  The emerging ratio is related to the ratio of the
Planck time $t_P$ and the age of the Universe $T$. 
\begin{equation}
N = f\left(\frac{T}{t_P}\right),.
\end{equation}
where $f$ is some growing function which should be close to exponential (supported by the chain character of branching) 
which leads to $N \sim10^{10^{60}}$. 
This number seems to be fairly huge in order to accommodate
all the unlikely events leading to the modern picture of Life, and there is no other reason at sight for 
such number appearance. 

Note that the relevance of the whole Universe for the life emergence makes natural the role of its distant
regions and life transmission by meteorites which is currently got a solid experimental support.

In fact, a similar opportunity (without these estimates) was explored by J. McFadden (McFadden, 2001) in the case of the earliest stage of
the biological evolution, where he expresesed a revolutionary idea that
the first life appears only in one of the innumerous Everett worlds.

However, the author dislikes the immediate consequences of his hypothesis which he absolutely correctly deduces: namely, that extraterrestrial life, and therefore, intelligence does not exist (note
the same hypothesis was suggested  for different reasons by I.S. Shklovsky who radically changed his earlier opinion (Shklovsky, 1968; Lem, 2001) and that the life cannot be created in the laboratory.

To overcome these obstacles he suggests the another use of quantum theory to explain the improbable event, namely, the inverse Zeno effect. However, we did not  consider this opportunity as a plausible one (Kamenshchik and Teryaev, 2008).

Indeed, he considers as a model of improbable event the passage of light through the
vertically and horizontally polarized lenses while the insertion of extra lenses
between them increases the probability.

This case, however, deals with low-dimensional system when the small probability
is achieved due to a sort of fine-tuning (mutual orthogonality of lenses).
At the same time, the low probability of a transition leading to a first self-replicator
is due to a large dimension of the corresponding Hilbert space. More quantitatively,
if one has two wave functions (normalized vectors in a Hilbert space) one of which $|i \rangle$,
corresponds to initial ``single amino acid arginine'' (McFadden,2000) while, the second, $|f \rangle$,
corresponds to the emerged self-replicator. The typical (average) value of the square of their scalar product (fidelity),  related to a transition probability is
\begin{equation}
< |\langle i | f \rangle|^2 > = \frac{1}{N},
\label{aver}
\end{equation}
where $N$ is a dimension of the Hilbert space defined by the number of participating elements.
Now, if one produces some quantum measurement, the scale of this quantity clearly remains the same.
The only way to increase these probabilities by a dense series of measurements would be to
arrange them in some particular way defined by the initial and the final states. The appearance of such a special measurement-like process
is not easier to explain than the occurrence of a small-probability quantum transition. At the same time, some random measurements will not substantially increase the probability (\ref{aver}), contrary to the case of polarized lenses, when the specially organized low probability may be increased by a generic measurement.

Therefore, we do not consider inverse quantum Zeno effect as a candidate for the explanation of low probability events necessary for life emergence and come back to the initial suggestion of McFadden
about the use of MWI.

\section{Mesoscopic Anthropic Principle and Biological Evolution}

While for the emergence of life there is the common agreement among the biologists that its current understanding is not 
satisfactory, it is not so for further evolution  
leading to the appearance of its complex forms.
The Darwinian evolution is an adaptive one (Dawkins, 1996) and explains the
arising of the complex structures if they provide the evolutionary advantages.
At the same time, the appearance of complex structures, which does not lead to immediate evolutionary success,
including the Human beings
is not trivial to explain. The production of complexity in the process of the type of
random walk may be explained (Gould, 1996) only if this complexity is relatively low. The random walk in that case is limited by zero complexity barrier and produces its increase. The further
evolutionary process explains the progress of most numerous species, like insects,
but not the appearance of  complex and rare ones. 

To explore the possible role of MWI in the evolution, we suggest to extend this mechanism to all the stages of biological evolution.
Indeed, the original suggestion of McFadden (2001) is to limit the field of applicability of
quantum effects to the microbilogical scale (McFadden, 1999) when the entanglement between
cell and its environment is essential, while for the multi-cell structures quantum effects
were considered (McFadden, 2001) unimportant.

Contrary to that, we suggest that {\it all} the mutations in the course of biological evolution are
the quantum measurent-like processes so that all their different outcomes are realized in different branches.
The increasing of complexity now has a purely random character, so that only in few parallel worlds
the biological evolution produces more and more complex species.

All the parallel worlds emerging due to mutations differ only by small variations in the
mutating organism. This feature is common with a standard (neo)Darwinian paradygm.
What is different from it is that {\it all} the versions of this variation are realized in different
parallel Everett worlds. This naturally implies the increase of complexity in some of them just by random process.
In our opinion, this solves the fundamental problem of the extremely law probability of life emergence and
evolution to the most complex forms, including ourselves.

There is a number of
fundamental facts which, to our opinion, do not contradict to or even support this hypothesis.
These are ``punctuated equilibrium'' (evolution proceeds by sudden bursts followed by
long ``stasis'' periods), ``Out of Africa'' theory (Leakey, 1996) (
appearance of all humans from a single family), ``Mitochondrial Eve''
(identifying a common female anchestor, being the support of previous theory),
``irreversibility of the brain formation'' (once emerged brain never reduced in the course of evolution) etc.

Another phenomenon which is currently considered as an evidence of a directed evolution (Markov, 2010)
is the parallelism observed before any major evolutionary advance. In MWI picture this may be explained 
as a collecting of improbable mutations in the Everett worlds selected by AP with the subsequent horizontal 
exchange of genes leading to the new evolutionary step.

It is interesting that MWI allows to describe the famous Gaia Theory (Lovelock, 2010) in completely Darwinian 
framework, as the critical remarks by R. Dawkins (Dawkins, 1982) on the necessity of the existing of multiple
competeng Gaia's may be indeed realized in parallel Everett worlds and the selection between them is performed by AP.

We have no opportunity of detailed discussions and just mention that all these facts
may be understood as emerging from improbable rare events of quantum measurement type,
so that all of their outcomes are realized in parallel worlds. We are just lucky inhabitants of one of the most ``pleasant'' of them.

Let us stress, however, that we consider the subject of this section as a more contradictory and 
problematic than that of previous one.
At the same time, its further investigation may open the striking opportunities to study MWI and, hopefully, even to falsify it. 

\section{Quantum Theory and Consciousness} 

The problem of possible interrelations between the consciousness and quantum mechanics 
has been attracting attention of researchers since the dawn of quantum mechanics. 
We would like to emphasize that this problem has three aspects. First, does quantum theory 
help to explain the origin of the consciousness ? Second, is the quantum theory necessary for 
the very existence of the consciousness ? Third, is the consciousness necessary for the very existence of quantum theory ? 

It seems to us that the ideas presented in the preceding sections answer, in a way, the first question. The combination of the many-worlds interpretation of quantum theory with the anthropic principle allows to explain the  biological evolution and its top result - the appearance of Human Mind. As far as the second question is concerned we inclined to think 
that the answer might be positive as well. Indeed, let us suppose that the world is described exclusively by the classical mechanics, and, hence, the ``Laplace demon'' does exist.
Namely, fixing exactly the coordinates and velocities of all the particles in the Universe, it 
is possible, in principle, to forecast its future evolution in all the tiny details. Thus, all the complexity of the universe is already contained in its initial state. How is it possible to prepare 
such a state?  Perhaps, the answer could be only highly teleological. In the framework 
of the many-worlds interpretation of quantum mechanics, the possibility of the co-existence 
of different alternatives, or Everett worlds, makes quite natural the fact that between this huge 
set of worlds does exist some small fraction of worlds, where the complex structures, including the Life and the Mind are present.

That the consciousness cannot emerge as a result of a ``classical'' computer operation, may be intitively seen by the 
observing of its mechanical version designed by Babbage and realized in London Museum of Science.      

The third question, if the consciousness is necessary to make the quantum mechanics self-consistent and to describe the process of quantum measurement, was discussed in the framework of the Copenhagen interpretation, in particular, by von Neumann in his book 
(von Neumann, 1955). As we have already mentioned, in the quantum measurement according to von Neumann, three actors played: the quantum object, the measuring device and the observer. This observer should have possessed the consciousness, otherwise 
the reduction cannot be completed. Here, it is necessary to mention that the description of this point given by von Neumann was a somewhat vague. He said that the borderlines between 
these three substances are not well defined, but only the fact of their existence is essential.

Another proponent of the Copenhagen interpretation, Niels Bohr, did not emphasize the role of consciousness in the process of quantum measurement and limited itself by using the notion of classical realm. On the other hand, Bohr used to trace the analogies between the complementarity in quantum physics and that in biology and psychology, and even found
something similar to the indeterminacy principle in the description of the process of thinking 
(Bohr, 1958). Thus, on the very qualitative level, Bohr speculated about the quantum nature 
of Life and Mind. Generally speaking, one can say that his answer on the second question was positive, but he did not insist on answering the third question.

The creation of the many-worlds interpretation  by Everett, was motivated by the desire to 
treat the quantum mechanics without using the classical realm or the preferred role of 
a conscious observer. All the quantum subsystems, participating in quantum interactions, 
including measurements and observations were treated on equal footing. It is known also
that Everett has dedicated essential efforts working in the development of an artificial 
intellect concepts (Byrne, 2010). Thus, one can believe that in a way he considered the human 
brain as a some kind of computer, which certainly, was governed by the laws of quantum mechanics, but was not an essential ingredient for its foundations. 

An interesting attempt of the combined answer on the second and third questions, formulated above, was undertaken by M.B. Mensky, who has elaborated the Extended Everett Concept
 (EEC) (Mensky, 2007a; Mensky, 2007b; Mensky, 2010; Mensky, 2011). He has stated that both questions should be answered positively. The consciousness cannot exist without the quantum
 theory, but the quantum theory cannot exist without the consciousness as well.  
The author of the (EEC) notices that the Everett interpretations looks complicated and not logical from the classical point of view, while it is very simple from the quantum point of view.
Indeed, one has only one fundamental process instead of two. However, the real difficulty 
arises when we speak about the separation of the alternatives. On the other hand nobody 
really knows what is the consciousness. Thus, Mensky proposes to resolve the problem of 
the absence of a good definition of these two phenomena by the identification between the consciousness and the splitting of classical realities. Thus, the emergence of parallel Everett worlds is an objective process 
(like the operation of particle detector at LHC)
and  consciousness is exactly what 
is responsible for the splitting of classical realities (when experimentalists sees only one of the possible outcomes of particles interactions). 

Coming back to the evolution, this splitting may be considered as a first major evolutionary
advance, as it is hardly possible to imagine the living creature operating simultaneously in the parallel worlds. 

However, the possibility of reaching 
of the parallel Everett branches do exist and can be realized in such particular states of 
consciousness as sleep or trance. The telepathy and other parapsychological effects can also find 
its explanation in the frame of the Extended Everett Concept. Besides, the consciousness can influence the quantum probabilities, augmenting the chances of realization of some 
events.  

Let us notice, that the idea that telepathy, telekinesis and other paranormal phenomena 
are in some way connected with the non-classical quantum-mechanical nature of our world 
has a relatively long history. A well-known writer and thinker Arthur Koestler was a proponent of this idea and devoted to it some papers and books, between each the most significant is 
``The Roots of Coincidence'' (Koestler, 1972). More recently, a Nobel Prize winner in physics B.D. Josephson has become an active supporter of this idea (Josephson and Pallikari-Viras, 1991).   One can easy understand how attractive is this prospective, but to our mind it is better to be more cautious. As far as we know, there are no testable quantum-mechanical models of  some parapsychological effects.

Coming back to the Extended Everett Concept, we can say that the intention to attract the quantum-mechanical methods to the study of the consciousness looks quite reasonable, but 
the indefiniteness of the process of splitting between alternatives and the absence of the understanding what the consciousness is, cannot be treated on equal footing. Indeed, 
as was explained in the third section of the present paper (see also the papers 
(Zeh, 1973; Barvinsky and Kamenshchik, 1990; Barvinsky and Kamenshchik, 1995a; 
Barvinsky and Kamenshchik, 1995b)). Here the subdivision of the quantum system into 
subsystems determines the preferred bi-orthogonal basis, whose elements are identified 
with different Everett worlds. These states are orthogonal according to their construction 
and there is no way in which the consciousness or some other agent can provide jumps 
between different branches or to influence the probabilities of different outcomes of experiments. On the other hand, the notion of the consciousness seems to us much more complicated than the definition of branches or Everett worlds (it is valid also for simplest forms 
of consciousness which could be identified with perception). Thus, the mutual definition of
the consciousness and of the world splitting through their identification does not look plausible.

All said above does not mean that the study of the interrelations between the quantum
 physics and the brain sciences is useless. Vice versa, we believe that one can expect 
 a new and an unpredictable development of the brain sciences, when the quantum informatics 
 would meet the physiology of brain, combined with some researches in the empirical psychology.      
  However, it is difficult to foresee now what concrete from of interaction between sciences 
  will bring us to success.

\section{Are Small Probabilities Irrelevant ?}

As is well known one of the main features of quantum mechanics is the 
probabilistic character of its predictions. During many years the Born rule, connecting the probabilities of the outcomes of a quantum measurement with the squared modules of the coefficients  of the expansion of the wave function of a system under consideration with respect
to the eigenvectors of the operator, representing the measured quantity, was considered 
as a fundamental postulate. However, later the so called Finkelstein-Hartle-Graham theorem 
was proven (Finkelstein, 1963; Hartle, 1968; Graham, 1973). The message of this theorem 
consists in the fact that the Born rule can be derived from other postulates of quantum mechanics if one considers a huge number of identical quantum systems and defines the probability as a relative frequency of a chosen outcome of an experiment with respect to 
the general number of trials. Finkelstein and Hartle considered this statement without connection with the many-worlds interpretation of quantum mechanics, while Graham 
worked in the framework of this interpretation. It seems to us that the Finkelstein-Hartle-Graham theorem looks especially harmonically in the framework of the many-worlds interpretation. Thus, for the convenience of the reader we shall present here a short demonstration of this theorem.    
 
Let us consider an experiment undertaken on $N$ identical quantum systems. Suppose that 
$M$ different outcomes of the measurement are possible. If $k$th outcome appears $m_k$ 
times, then the relative frequency of this outcome is equal to $\frac{m_k}{N}$; naturally,
$\sum_{k=1}^{M} m_{k} = N$. 

The probability of $k$th outcome can be measured as 
\begin{equation}
\lim_{N \rightarrow \infty} \frac{m_k}{N}.
\label{frequency}
\end{equation}
Commenting Eq. (\ref{frequency}) we have used the term ``measured'' instead of 
``defined'' . The point is that the term ``defined'' sometimes is used in the sense that 
the very notion of probability is identical to the notion of relative frequency and does not 
have another sense (von Mises, 1964). We postpone a little bit the discussion of this question 
and shall limit ourselves by the remark that the formula (\ref{frequency}) gives an operational method of calculation of the probability. 

Let us consider a quantum  system, described by the wave function 
\begin{equation}
|\Psi \rangle = \sum_{i=1}^{M}C_i |\psi_i \rangle,
\label{psi}
\end{equation}   
where $|\psi_i \rangle$ are the eigenstates of the operator $\hat{A}$, corresponding 
to its eigenvalues $A_i$. Let us consider now a set of $N$ identical systems, described 
by the wave function $|\Psi \rangle$. Such a set is described by the wave function 
\begin{equation}
\underbrace{|\Psi \rangle\otimes\cdots\otimes|\Psi \rangle}_{N\ {\rm times}}.
\label{product}
\end{equation} 

Now we introduce the operator of the relative frequency $\hat{F}_N^k$, indicating the relative frequency for the outcome $A_k$ in the series of $N$ experiments. Then
\begin{eqnarray}
&&\hat{F}_N^k\underbrace{|\psi_1\rangle\otimes\cdots\otimes|\psi_1\rangle}_{m_1\ {\rm times}}\cdots\underbrace{|\psi_k\rangle\otimes\cdots\otimes|\psi_1\rangle}_{m_k\ {\rm times}}
\cdots\underbrace{|\psi_M\rangle\otimes\cdots\otimes|\psi_M\rangle}_{m_M\ {\rm times}}\nonumber \\
&&=\frac{m_k}{N}\underbrace{|\psi_1\rangle\otimes\cdots\otimes|\psi_1\rangle}_{m_1\ {\rm times}}\cdots\underbrace{|\psi_k\rangle\otimes\cdots\otimes|\psi_1\rangle}_{m_k\ {\rm times}}
\cdots\underbrace{|\psi_M\rangle\otimes\cdots\otimes|\psi_M\rangle}_{m_M\ {\rm times}},
\label{freq1}
\end{eqnarray}
where $\sum_{i=1}^{M} = N$. 

Now, we should prove that at $N \rightarrow \infty$ the relative frequency of the outcome 
$A_k$ tends to $|C_k|^2 = p_k$. To do it let us consider the vector 
\begin{equation}
(\hat{F}_N^k-p_k)\underbrace{|\Psi \rangle\otimes\cdots\otimes|\Psi \rangle}_{N\ {\rm times}}
\label{freq2}
\end{equation}
and show that its norm tends to zero at $N \rightarrow \infty$.  Then, the positivity of the scalar product on the Hilbert space implies the tending of the vector (\ref{freq2}) to zero.
Thus, in the limit $N \rightarrow \infty$ the wave function (\ref{product}) becomes the eigenfunction of the operator 
of relative frequency with the eigenvalue $p_k = |C_k|^2$. 

Now, substituting (\ref{psi}) into (\ref{freq2}), we can write down the squared norm of the 
latter:
\begin{eqnarray*}
&&\underbrace{\langle\Psi|\otimes\cdots\otimes\langle\Psi|}_{N\ {\rm times}}(\hat{F}_{N}^{k}-p_k)^2\underbrace{|\Psi\rangle\otimes\cdots\otimes|\Psi\rangle}_{N\ {\rm times}} \nonumber \\
&&=\sum_{\stackrel{m_j=0}{\sum_{j=1}^{M}m_j=N}}^{N}|C_1|^{2m_1}\cdots |C_k|^{2m_k}\cdots |C_M|^{2m_M}\frac{m_k^2}{N^2}\frac{N!}{m_1!\cdots m_M!}\nonumber \\
&&-2p_k\sum_{\stackrel{m_j=0}{\sum_{j=1}^{M}m_j=N}}^{N}|C_1|^{2m_1}\cdots |C_k|^{2m_k}\cdots |C_M|^{2m_M}\frac{m_k}{N}\frac{N!}{m_1!\cdots m_M!} + p_k^2\nonumber \\
\end{eqnarray*}
\begin{eqnarray}
&&=\sum_{m_k=0}^{N}\frac{m_k^2}{N^2}\frac{N!}{m_k!(N-m_k)!}
\sum_{\stackrel{m_j=1}{\sum_{\stackrel{j=1}{j\neq k}}^{M}m_j=N-m_k}}^{N}
\frac{(N-m_k)!}{m_1!\cdots m_M!}|C_1|^{2m_1}\cdots |C_M|^{2m_M}+p_k^2\nonumber \\
&&=\sum_{m_k=0}^{N}\left(\frac{m_k^2}{N^2}p_k^m(1-p_k)^{N-m_k}\frac{N!}{m_k!(N-m_k)!}-\frac{2m_k}{N}p_k^{m_k+1}(1-p_k)^{N-m_k}\frac{N!}{m_k!(N-m_k)!}\right)\nonumber \\
&&+p_k^2=p_k^2\left(\frac{N-1}{N}+\frac{1}{Np_k}-2+1\right)=\frac{p_k(1-p_k)}{N}.
\label{freq3}
\end{eqnarray}
When $N \rightarrow \infty$ the expression (\ref{freq3}) tends to zero and  the theorem is proven.  

Thus, the Born rule can be derived from the consideration of the wave function of the system, including an infinite (in the limit) number of identical subsystems. The probability is in a natural way connected with the scalar product on the Hilbert space. 

However, in the treatment of the notion of probablity in the frame of the many-worlds interpretation of quantum mechanics 
some problems arise. One of them can be called the problem of two measures. As was noticed by Graham (1973) in the many-worlds interpreation 
another measure can be introduced. This measure is connected with a naive counting of the Everett worlds in which different outcomes of 
an experiment are realized. From the point of view of usual quantum-mechanical measure these worlds do not have the same value because to any of them corresponds some probability $p_i = |C_i|^2$. 

However, one can think in another way (Graham, 1973). One can say that from the point of view of the many-worlds interpretation all 
these worlds are equally real and we cannot prescribe some weights to them. If we would like to know the probability of a certain concrete 
event, we should consider the set of all Everett worlds and then to learn in which part of them the given event occurs. Thus, the new
``naive'' probability can be defined as 
\begin{equation}
\tilde{p} = \frac{m}{N}, 
\label{naive}
\end{equation}
where $N$ is a full number of the Everett worlds and $m$ is number of the worlds where the event which we are interested in occurs.
Obviously, the probabilities defined in (\ref{naive}) can differ essentially from the usual quantum-mechanical probabilities.

To make the conception more clear, let us consider an example. We have an object described by the wave function 
\begin{equation}
|\Psi\rangle = C_1|\Psi_1\rangle + C_2|\Psi_2\rangle.
\label{2-meas}
\end{equation}
Let us immagine an experiment with two outcomes correposnding to the functions $|\Psi_1\rangle$ and $|\Psi_2\rangle$ 
with the probablities $p = |C_1|^2$ and $1-p = |C_2|^2$. Having done the experiment on the 10 identical systems, we shall see that the wavefunction describing these ensemble toghether with an observer will be splitted into $2^{10} = 1024$ branches. Let us calculate now the probability of getting of the same result corresponding to the wave function $|\Psi_1\rangle$ in all 10 experiments. 
Obviously this occurs only in one world.
According to the Graham definition(\ref{naive}) we have the probablity of such an event equal to $2^{-10}$ while according to 
the standard Born rule the probability is equal to $p^{10}$ and can be quite big if $p$ is close to 1. Continuing the comparison between 
these two measures, one can see that the most probable from the Born point of view distribution of outcomes 
(approximately in $10p$ cases the first outcome and in $10(1-p)$ cases the second outcome) will be realized only in a small part of 1024 
Everett worlds. In the majority of worlds one shall see 5 first outcomes and 5 second outcomes. 

In other words, only in a small subset of the Everett worlds the usual predictions of the quantum mechanics will be confirmed. 
However, if we take into account the weights of these branches according to the Born rule, then the weighted part of the worlds, 
where the quantum mechanical predictions are false, will be very small, because such  worlds enter into the expansion of the wave function with very small coefficients.

Thus, we have seen the contradictions between two measures, but which of them is true ? The quantum mechanical measure is supported by the mathematical apparat of quantum mechanics, including the Finkelstein-Hartle-Graham theorem and by the experimental data. The naive measure 
is connected with the idea of the equally real parallel Everett worlds. If the quantum mechanical measure is true and the naive probablity does not have sense, does not it compromize the many-worlds interpretation?

Curious, while to our mind wrong, attempt of resolution of this contradiction was undertaken by Graham (1973). His approach was critisized
in the book (Barvinsky, Kamenshchik and Ponomariov, 1988) and  our exposition follows here to that represented in this book. 
Graham considered the naive probablity as a fundamental one, and using thermodynamical methods, has tried to reduce the quantum mechanical measure to the naive one. The Graham's logic was the following one: the quantum measurement is done by a macroscopic device, thus it includes 
such a stage as a transition of the disturbed device into a state of thermal equilibrium. To the macrostate of the device corresponds a whole
subspace in the Hilbert space of states and one can think that all the microstates corresponding to a given macrostate have the same weight.
Starting from these correct statements Graham suggested that the process of the transition of the device to the thermal equilibrium is 
combined with the process of some kind of  statistical smoothing of the wave function of the quantum object. To describe this process Graham has undertaken the averaging of the wave function of the object under consideration on all the Hilbert space. Then one can find the average values of the squared modules of the coefficients of the expansion of the wave function of the object on the basis, including the states corresponding to the definite outcomes of the experiment. These average values are equal to $N^{-1}$, where $N$ is the number of possible outcomes of the experiment (for details see (Graham, 1973) and 
(Barvinsky, Kamenshchik and Ponomariov, 1988)). Having done these direct calculations, Graham has concluded that using the 
thermodynamical notions he had managed to reduce the quantum mechanical measure to the naive one. 

However, in this averaging  the principal error is hidden. This procedure eliminates all the quantum information, encoded in the 
coefficients of the expansion of the wave function. Thus, this procedure is equivalent to the obviously wrong hypothesis that all the coefficients of the expansion are equal. 

We have discussed this attempt of the reducing of  the Born probability to the naive probability to show that the difference 
between them are fundamental and one of them cannot be reduced to another. Then, what is the sense of the probability in the many-worlds interpretation when all the outcomes are realized ? From our point of view, the natural way of interpretation of 
the probability in quantum mechanics in general and in the many-worlds interpretation, in particular, is the acception of the Popper's conception of the probability as a ``propensity'' (Popper, 1959). 

Probably, the most known interpretation of the notion of probablity (at least, between physicists) is relative-frequency interpretation, elaborated by von Mises (von Mises, 1964). According to this interpretation the relative frequency 
measurement is not only  the empirical method of the calculation of the probability, but the definition of the very notion 
of the probability. Thus, the probability is applicable only to the study of series of trials or of ensembles of identical objects and is not applicable to individual events and objects. At the same time, the probablistic character of the physical laws is fundamental according to von Mises. The applicability of the Mises approach to the statistical elboration of data or 
to the classical statistical physics does not provoke doubts. The relative-frequency conception of the probability 
 enters naturally into the statistical interpretation of quantum mechanics, which rejects the applicability of the 
quantum mechanics to individual systems, limiting itself  by consideration of ensembles (see e.g. the book 
(Blokhintsev, 2010)). 
However, already the Copenhagen interpretation considers individual objects and hence, here the probablities are not subject 
to the relative-frequency treatment. Analysing this problem Popper has suggested the conception of the probablity as a propensity (Popper, 1959). He considered the probability as a property of the event by itself and not as a characteristic 
of a succession of events or of an ensemble of objects. One can say that the probability treated as a propensity is a quantitative measure of tendencies, which an object under consideration possesses. The propensitive interpretation of the probability gives us an opportunity to speak about the probability of an individual event, Thus, we can apply quantum mechanics to individual objects being free of rather narrow frames of the statistical interpretation.

The probability as a propensity becomes especially attractive when we combine it with the many-worlds interpretation of quantum mechanics. Indeed, the probability coefficients, describing the different branches of the wave function, characterize the propensities  of an observer to come to one of the parallel Everett worlds, while the relative-frequency interpretation of the probability is hardly acceptable in this context. Indeed, the act of branching, or the defactorization of the wave function is unique and one cannot consider a statistical ensemble of identical universes, in each of which the branching takes place.  

Now, let us note that when one speaks about the universe as a whole, combining the many-worlds interpretation with the anthropic
principle, even the propensity interpretation of the probability has a relative value. Rather often, the reaserchers 
trying to explain why, for example, a cosmological constant or other important characteristics of the universe have values favorable for the appearance of the Life and of the Mind, make some estimations of the probabilistic distributions for these 
characteristics. From our point of veiw it hardly can have some sense. The branches of the wave function of the universe, where 
these favorable values are realised have a very small probability weights. It is not important. It is important that such branches do exist ! Thus, at least, in a small part of the Hilbert space some interesting development is possible.
Once again, the comparison between small and large probabilities makes sense when operates with large numbers of identical systems or with the large number of identical experiments. When one thinks about unique events in the history of the universe, 
the real difference is between possible and impossible, in principle.

It is interesting to note that the understanding of the fact that the events which have very small probability can represent an especial interest  was recently developed in the field, 
which looks to be non connected with physics, namely in the social life, including its political 
and economical aspects. We mean here the bestselling book ``The Black Swan'' by  
N.N. Taleb, a modern expert in finances and statistics and a philosopher (Taleb, 2007).
The main topic of the book is the analysis of the role of events which looks extremely 
improbable and unpredictable, which the author calls ``black swans''. Paradoxically, 
these events rather often appear to be the most significant for the future development 
of the Humankind. The appearance of such events and their influence is typical for very 
complicated systems. 

Another interesting observation concerning the qualitative importance of quantitatively 
small objects one can find in the article ``The Birth of the Freedom'', written by a Russian    
philosopher Georgy Fedotov as early as in 1944 (Fedotov, 1944). The translation of some 
extractions from this text is worth of citing.

``The puzzle of the importance of small magnitudes remains unsolved: why almost all which 
has a great value occurs at the materially small scales? \ldots We should upside down all 
the scales of estimations and take into account as an initial point not the quantities, but the qualities.\ldots The freedom shares the destiny of all elevated and valuable in the world.
A small, politically fragmented Greece has given to the world the science, has given the forms of the thought and of the artistic perception which until now determine the world view of 
hundreds of millions of persons. Even more tiny Judea has given to the world the greatest 
religion, which is professed by people of all the continents. The small island behind 
La Manche has elaborated the system of political organization, which being less universal than the Christianity and science, reigns, nevertheless in three continents.''

Coming back to the physics, we can say that the most important branches of the universe,
or the most interesting Everett worlds can have a very small probabilistic weight.  Thus, like in the Taleb's statements that the most 
important events are unpredictable from the usual probabilistic point of view, or in the 
Fedotov's observation that the most important developments in the history of the Humankind 
were realized in small geographical and politically weak areas, we can say that the most 
interesting things in the universe occur in a tiny part of the Hilbert space, where the wave function of the universe lives.

\section{Discussion and Conclusions}

In this article we have tried to explore the possible relation between Anthropic Principle and
Many-World Interpretation of Quantum Theory. The key moment is the
possibility to multiply the reality to such an extent that very special events like emergence of Life become quite possible.

The important feature of this process is the smallness of differences between various parallel Everett worlds. This allows to scan all the possible values of required parameters which is essentially
similar to the arguments justifying Darwinian natural selection. The only, albeit crucial difference is
that selection occurs not in the different moments of time like Darwinian one, but in
the different parallel worlds, or, mathematically speaking,
in the different regions of Hilbert space.
Such a resemblance to the Darwinian evolution may be explored
for other known mechanisms of generation of variety of options (like string landscape
or eternal chaotic inflation) in order to separate the "physical" predictions from
the effects of ``environment'' (Rubakov, 2006) or ``scanning'' (Weinberg, 2005) which we are
about to suggest.

Indeed, if some physical constant should be fine-tuned for the emergence of life it
is very unlikely that it is completely defined by underlying physics (cf. (Smolin, 1998))
and selection process of Darwinian type was likely to contribute.
At the same time, the physics should rather lead to the establishing of general framework
and more robust constraints (see, for example, (Barvinsky and Kamenshchik, 2006a;  Barvinsky and Kamenshchik, 2006b), where
in the framework of the Euclidean quantum gravity some constraints
on possible values of the effective cosmological constant were found) which
may be a starting point for subsequent fine-tuning by anthropic selection.

In the case of the Many-Worlds Interpretation such a selection allows to fine-tune
various parameters which are not amongst the basic constants of theory of fundamental interactions,
including gravity and elementary particle physics. This is because the branching due to the
Many-Worlds interpretation occurs when all the fundamental constants are already fixed and
therefore they are the same in all the Everett parallel worlds.
We suggested to use the term ``Mesoscopic Anthropic Principle'' for the description of
anthropic selection in the branching process.

We considered two possible fields of applicability of Mesoscopic Anthropic Principle,
namely, planetary coincidences and biological evolution.

In both cases the small differences generated by branching allow to explain the coincidences which is very difficult to do otherwise. As an example  we consider the coincidence of angular sizes of Sun and Moon
responsible for the Solar eclipses. This coincidence may be achieved by small steps during branching,
and anthropic selection may choose it to be realized in our Universe if eclipses played any role
in the life emergence. This hypothesis may be checked , in principle, opening an opportunity for indirect tests
of Anthropic Principle.

The other important problem is the arising of complexity during bilogical evolution,
 including such extreme cases as Life itself and Mind. We suggest that crucial role is played
the Many-Worlds interpretation,
so that extremely small probability is fully compensated by enormous number of trials.

Summing up, we consider the Anthropic Principle combined with the multiple opportunities
opened also by the Many-Worlds interpretation of quantum theory, as new exciting field
of physics and other natural sciences, rather than dull alternative to them.

\section*{Acknowledgment}
This work was partially supported by the RFBR Grant No 11-02-00643. 

\section*{References}
Albrecht A. Investigating decoherence in a simple system. Phys Rev D 1992; 46: 5504-5520. 
\\  
Artru X, Elchikh M, Richard J-M, Soffer J and  Teryaev O.
Spin observables and spin structure functions: inequalities and
dynamics. Phys Rept 2009; 470: 1-92.   \\
Barrow JD and Tipler FJ. The Anthropic Cosmological Principle. Oxford University Press 1988.\\ 
Barvinsky AO, Kamenshchik AY and Ponomariov VN. Fundamental Questions of the Interpretation of Quantum Mechanics, A Modern Approach. Publishing House of the
Moscow Pedagogical University 1988.\\
Barvinsky AO, Kamenshchik AY and Ponomariov VN. Anthropic Principle and Many-Worlds Interpretation of Quantum Mechanics. In Proceedings of the International Seminar  
``Anthropic Principle in the Structure of  Scientific Picture of the World'', November  28-30,
1989, Leningrad, 48-50.\\
Barvinsky AO and Kamenshchik  AY. Preferred basis in the
many-worlds interpretation of quantum mechanics and quantum
cosmology. Class Quantum Grav 1990; 7: 2285-2293.\\ 
Barvinsky AO and Kamenshchik  AY.
Preferred
basis in quantum theory and the problem of classicalization of
the quantum Universe. Phys Rev D  1995a; 52: 743-757.\\
Barvinsky  AO and  Kamenshchik  AY. Preferred basis in the
many-worlds interpretation in quantum theory and the symmetries
of the system. Grav Cosmol 1995b; 1: 261-265.\\
Barvinsky AO and Kamenshchik AY. Cosmological landscape
from nothing: Some like it hot. JCAP 2006a; 0609: 014.\\ 
Barvinsky AO and Kamenshchik AY.
Thermodynamics
via Creation from Nothing: Limiting the Cosmological
Constant Landscape. Phys Rev D  2006b; 74: 121502(R).\\
Ben Dov Y. An observer decomposition for Everett's theory. Found Phys Lett 1990; 3: 383-387.\\
Bioy Casares A. La trama celeste. Sur, Buenos Aires 1940.\\ 
Blokhintsev DI. The Philosophy of Quantum Mechanics. Springer, Berlin Heidelberg 2010.\\
Bohm D and Hiley BJ. The undivided universe : an ontological interpretation of quantum theory. Routledgge, London 1993.\\
Bohr N. Atomic physics and human knowledge. John Wiley and Sons, New York 1958.\\
Borges JL. El jardin de senderos que se bifurcan. Sur, Buenos Aires 1941.\\
Born M. Zur Quantenmechanik der Sto\ss vorg\"ange. Zeit Phys 1926; 37(12): 863-867.\\
Born M. Physics in my generation : a selection of papers. Pergamon Press, London 1956.\\
Byrne P. The Many Worlds of Hugh Everett III: Multiple Universes, Mutual Assured Destruction, and the Meltdown of a Nuclear Family,
Oxford University Press 2010.\\
Carter B. Large Numbers in Astrophysics and Cosmology. Paper presented at Clifford Centennial Meeting, Princeton 1970.\\
Chernavsky D.S., Synergetics and Information, Moscow 2009. \\
Dawkins R., The Extended Phenotype. Oxford University Press 1982.\\
Dawkins R. Climbing Mount Improbable. Penguin, London
1996.\\
Dawkins R., The God delusion. Penguin, London 2009.\\
Deutsch D. Quantum theory as a universal physical theory. Int J Theor Phys 1985; 24: 1-41.\\
DeWitt BS. Quantum mechanics and reality. Physics Today 1970; 23 (9): 30-35.\\ 
DeWitt BS and Graham N (Eds.). The Many-Worlds Interpretation of Quantum Mechanics.
Princeton University Press 1973.\\
Dicke RH. Dirac's Cosmology and Mach's Principle. Nature 1961; 192: 440-441.\\
Dieks D.  Resolution of the measurement problem through decoherence of the quantum state.
Phys Lett A 1989; 142: 439-446.\\ 
Dirac PAM. The cosmological constant. Nature 1937; 139: 323.\\
Dirac PAM. New basis for cosmology. Proc R Soc A 1938; 165: 199-208.\\
Efremov AV and Teryaev OV. On High P(T) Vector Mesons
Spin Alignment. Sov J Nucl Phys 1982; 36: 557.\\
Everett H. Relative-state formulation of quantum mechanics. Rev Mod Phys 1957; 29: 454-462.\\
Fedotov GP. Rozhdenie svobody. Novyj zhurnal, New York 1944; 8: 198-218.\\
Finkelstein D. The logic of quantum physics. Trans. N.Y. Acad. Sci. 1963; 25: 621-637.\\
Garriga J, Linde A and Vilenkin A. Dark energy equation of state and anthropic selection.
Phys Rev D 2004; 69: 063521.\\
Garriga J, Schwartz-Perlov D, Vilenkin A and Winitzki S. Probabilities in the inflationary multiverse. JCAP 2006; 0601: 017. \\
Giulini D, Joos E, Kiefer C, Kupsch J, Stamatescu I-O and Zeh HD. Decoherence and the Appearence of a Classical World in Quantum Theory. Springer-Verlag, Berlin-Heidelberg 1996.\\ 
Gonzalez G. and Richards JW. The Privileged Planet. Regnery
Publishing, Washington D.C. 2004.\\
Gould SJ. Life's Grandeur: The Spread of Excellence from
Plato to Darwin. Trafalgar Square, Cape 1996.\\
Graham RN. The measurement of relative frequency. In DeWitt and Graham (Eds.) 1973:
229-252.\\
Hawking S. Is the End in Sight for Theoretical Physics?
In Black Holes and Baby Universes and other essays. Bantam Press, London 1993.\\
Hartle J. Quantum Mechanics of Individual Systems. Amer. J. Phys. 1968; 36(8): 704-712.\\
Heisenberg W. \"Uber den anschauliehen Inhalt der quantentheorischen Kinemaitik und Mechanik. Zeit. Phys. 1927; 43: 172-198.\\
Heisenberg W. Physics and Philosophy: The Revolution in Modern Science. Harper and Brothers, New York 1958.\\
Ivnev R. Vladivostokskij starik. Moscow 1927.\\
Jammer M. The philosophy of quantum mechanics : the interpretations of quantum mechanics in historical perspective.
John Wiley and Sons, New York 1974.\\  
Josephson BD and Pallicari-Viras F. Biological Utilisation of Quantum NonLocality.
Found Phys 1991; 21: 197-207.\\
Kamenshchik AY and Teryaev OV. Many-worlds interpretation of quantum theory and 
mesoscopic anthropic principle. Concepts of Physics 2008; V(4): 575-592. arXiv:0705.2494 [quant-ph].\\
Koestler A. The Roots of Coincidence. Random House, New York 1972.\\
Leakey R. The Origin of Humankind Basic Books, New York
1994.\\
Lem S. Repetycja. Tygodnik Powszechny 2010, No 34.\\
Lovelock, J. Vanishing face of Gaia, Penguib Books, 2010, p. 111.\\
Linde AD. Particle Physics and Inflationary Cosmology. Harwood, Chur, Switzerland 1990.\\
Markov A., The Birth of Complexity, Moscow,2010 (in Russian).\\
 Markov MA and Mukhanov VF. Classical preferable basis in quantum mechanics. Phys Lett A 
 1988; 127: 251-254.\\
McFadden J and  Al-Khalili J. A quantum mechanical model of
adaptive mutation. BioSystems  1999; 50: 203-211.\\
McFadden J. Quantum Evolution. W. W. Norton and Company,
New York 2001.\\
Mensky MB. Postcorrection and mathematical model of life in Extended Everett's Concept. 
NeuroQuantology 2007a; 5(4): 363-376.\\
Mensky MB. Quantum measurement, the phenomenon of life, and time arrow: Three great problems of physics (in Ginzburg's terminology) and their interpretation. Phys Usp 2007b; 50: 397-407.\\
Mensky MB. Consciousness and Quantum Mechanics: Life in Parallel Worlds (Miracles of Consciousness from quantum Mechanics). World Scientific, Singapore 2010.\\
Mensky MB. Mathematical Models of Subjective Preferences in Quantum Concept of Consciousness. NeuroQuantology 2011; 9(4): 614-620.\\
von Mises R. Mathematical Theory of Probability and Statistics, Academic Press, New York 1964.\\
Muller B. The anthropic principle revisited. astro-ph/
0108259 2001.\\
von Neumann J. Mathematical Foundations of Quantum Mechanics.
Princeton University Press 1955.\\
Nielsen MA and Huang IL. Quantum Computation and Quantum Information. Cambridge University Press 2000.\\
Pogosian L, Vilenkin A and Tegmark M. Anthropic predictions for vacuum energy and neutrino masses. JCAP 2004; 0407: 005.\\
Panov A.D., Probabilistic interpretation of anthropic principle and Multiverse: in{\it Modern cosmology: Philosophical horizons}, Moscow, 2011, p.270-293 (in Russian). \\ 
Pohl F. The Coming of the Quantum Cats. Bantam Books, New York 1986.\\
Popper KR. The propensity interpretation of probability. 
The Brit. J. Phil. Soc. 1959; 10 (37): 25-42.\\ 
Prigogine I. From being to becoming: time and complexity in the physical sciences. Freeman and co. San Francisco 1980.\\
Rozental IL. Physical laws and numerical values of fundamental constants. Sov Phys Usp 
1980; 23: 296-305.\\   
Rozental IL. How Particles and Fields Drive Cosmic Evolution. Springer, Berlin 1988.\\
Rozental IL. Elementary particles and cosmology (metagalaxy and universe). Phys Usp
 1997; 40: 763-772.\\
Rubakov VA. Talk at XXXIII International conference on high energy physics, Moscow,
July 26 - August 2, 2006.\\
Schr\"odinger E. Discussion of probability relations between separated systems. 
Proc Cambridge Philos Soc 1935; 31: 555-563.\\ 
Schr\"odinger E. Probability relations between separated systems. 
Proc Cambridge Philos Soc 1936; 32: 446-452.\\
Shimony A. Degree of entanglement. Ann N Y Acad Sci 1995; 755 :
675.\\
Shklovsky IS and Sagan C. Intelligent Life in the Universe.
Holden-Day, San Francisco 1968.\\
Smolin L. Life of the Cosmos. Oxford University Press 1998.\\ 
Susskind L. The Anthropic landscape of string theory. arXiv: hep-th/0302219 2003.\\
Taleb NN. The Black Swan. Random House, New York 2007.\\
Teryaev OV. T odd effects in QCD. RIKEN Rev 2000; 28: 101-104.\\
Teryaev OV. The irreversibility of QCD evolution equations.
Phys Part Nucl 2005; 36: S160-S163.\\
Vilenkin A. Many Worlds in One. The Search for Other Universes. 
MacMillan, New York 2006.\\
Vilenkin A. Private communication 2010.\\
Weinberg S. Living in the multiverse. arXiv: hep-th/0511037 2005.\\ 
Zaslavsky GM. Chaos, fractional kinetics, and anomalous
transport. Phys Rept 2002;  371: 461-580.\\
Zeh HD. On the Interpretation of Measurement in Quantum Theory.
Found Phys 1970; 1: 69-76.\\
Zeh HD. Towards a quantum theory of observation. Found Phys 1973; 3: 109-116.\\
Zurek WH. Pointer basis of quantum apparatus: Into what mixture does the wave packet collapse ? Phys. Rev. D 1981; 24: 1516-1525.\\
Zurek WH. Environment-induced superselection rules. Phys. Rev. D 1982; 26: 1862-1880.\\ 

\end{document}